\documentstyle[prl,aps]{revtex}
\begin{document}
\draft
\title{Scaling with respect to disorder in time-to-failure}
\author{Didier Sornette$^{1,2}$
and  J\o rgen Vitting Andersen$^3$} 
\address{$^1$ Laboratoire de Physique de la Mati\`{e}re Condens\'{e}e, CNRS
URA 190\\
Universit\'{e} de Nice-Sophia Antipolis, Parc Valrose, 06108 Nice, France}
\address{$^2$ Department of Earth and Space Sciences and Institute of
Geophysics and Planetary Physics\\ University of California, Los Angeles,
California
90095-1567}
\address{$^3$ 
Department of Mathematics,
Imperial College,
Huxley Building, \\ 
180 Queen's Gate, London SW7 2BZ, England}

\date{\today}
\maketitle
\begin{abstract}
We revisit a simple dynamical model of rupture in random media with long-range
elasticity to test whether rupture can be seen as a first-order or a critical
transition. We find a clear scaling of the macroscopic modulus as a function of
time-to-rupture and of the amplitude of the disorder, which allows us to collapse
neatly the numerical simulations over more than five decades in time and more than one decade
in disorder amplitude onto a single master curve. We thus conclude that, at least in 
this model, dynamical rupture in systems with long-range elasticity is
a genuine critical phenomenon occurring as soon as the disorder is non-vanishing.
\end{abstract}
\vskip 0.5cm


\section{Small versus large heterogeneity in systems with limited load transfer}

Materials break down according to two broadly defined scenarios.
In the first one, examplified by a pure crystal, there is no or little damage
up to the rupture which occurs suddenly with no appreciable
precursors. In the second scenario that applies 
ideally in the limit of very heterogeneous media, the system is progressively 
damaged, first in an uncorrelated way reflecting the pre-existing
heterogeneity. As stress or strain increases, the damage becomes 
more and more correlated with crack growth and fusion, announcing the
incipient rupture. This second regime is like percolation at the beginning
and correlated percolation later and at the end of the process. In
the limit of infinite disorder, the rupture (in the quasi-static
limit) can actually be mapped exactly onto a percolation problem \cite{RouxHansen}.
This second scenario is characterized by a growing 
susceptibility and well-defined precursors. This classification 
has been emphasized by Mogi \cite{Mogi} in his search 
for earthquake precursory phenomena. Basing his reasoning on 
an analogy between elastic shocks (acoustic emissions) caused by fracture in 
heterogeneous materials \cite{AE} and earthquakes, he
noticed that the fracture process strongly depends on the degree of
heterogeneity of materials\,: the more heterogeneous, the more warnings
one gets; the more perfect, the more treacherous is the rupture. The
failure of perfect crystals is thus unpredictable while the fracture of
dirty and deteriorated materials could be forecasted. For once, complex
systems could be simpler to apprehend\,! However, since its
inception, this idea has not been much developed because it is hard to
quantify the degrees of ``useful'' heterogeneity, which probably depend
on other factors such as the nature of the stress field and boundary conditions,
the presence of
chemical contaminants, etc. This analogy nevertheless led Mogi
to hope that earthquake precursors could be identified for forecasting.
Nowadays, finding clear earthquake precursors is an active and
controversial research domain whose ultimate objective is still eluding 
the scientific community \cite{Kagan}. 

In contrast, the situation is more favorable in the
laboratory and in models. For systems where the load transfer has
limited stress amplification, the situation has been clarified.
By limited stress amplification, we refer to the cases where the stress
transfer due to a crack or more generally a damage zone does not exhibit
spatial concentration. This can occur in a variety of ways, for instance, 
in the democratic fiber bundle model \cite{DFBM} with the democratic rule
of stress transfer, or in 
models of block-springs with stick-slip behavior induced by solid friction in
which the stress transfer is screened beyond the 
sticking blocks \cite{Langer}. In real systems, fiber-matrix composites exhibit this
property as a local fiber rupture is locally accomodated by a distorsion of
the matrix which screens any local concentration and help uniformize the 
stress \cite{Fibers}. For such systems, it has been
shown recently \cite{First} that
disorder is a relevant field leading to tri--criticality, separating a
first-order (or abrupt) regime where rupture occurs without significant precursors from a
second-order (or continuous critical) 
regime where the macroscopic elastic coefficient exhibit power law
behavior. These results have been obtained using analytical solutions of fiber bundle models
and numerical simulations of a two-dimensional tensorial spring-block system in
which stick-slip motion and fracture compete \cite{First}. The idea is that,
upon loading a heterogeneous material, single isolated microcracks appear
and then, with the increase of load or time of loading, they both grow and
multiply leading to an increase of the number of cracks.
As a consequence, microcracks begin to merge until a ``critical density''
of cracks is reached at which the main fracture is formed. It is
then expected that various physical quantities (acoustic emission, elastic,
transport, electric properties, etc.) will vary. However, the nature of
this variation depends on the heterogeneity. The new classification uses the
fact that there
is a threshold that can be calculated\,: if disorder is too small, then the
precursory signals are essentially absent and prediction is impossible. 
In the language of phase transitions, the heterogeneity
is a control parameter (like the chemical potential in the Blume-Emery-Griffith
Ising model) that controls the distance to a so-called tri-critical
transition as the disorder increases, from a Griffith-type
{\footnote{The Griffith criterion for rupture takes exactly the form
of a condition for a critical ``droplet'' to nucleate and trigger
the growth of the new phase \cite{Gunton}.}} abrupt rupture
(``first-order'') regime to a progressive damage ending at
rupture, corresponding to a critical or ``second-order'' transition.
The other control parameter controls the distance to global rupture and can
be time, strain, stress or combinations of these.
The transition between the two regimes, which are 
two modes of brittleness culminating in a sudden failure, is different from
the brittle-ductile or brittle-plastic transitions. The value
of the disorder threshold separating these two regimes depends on the
system strength and other properties. This has in fact been tested
carefully in several laboratory rupture experiments, for instance using
acoustic emissions as precursors of rupture of fiber composites\cite{Anifrani}. The idea
that rupture can be ``critical'' is not new \cite{critical} but the classification
of why, when and how much so is useful.

\section{The case of long-range elasticity}

The purpose of this note is to complement these results by studying a case 
where the stress load transfer has {\it no} limiting amplification in amplitude
and range, as for instance in genuine elasticity. In this case, a fracture of length $a$
produces a stress amplification factor proportional to $\sqrt{a}$ 
extending to large distances, typically with a
$r^{-d}$ decay in $d$ dimensions for $r \gg a$, as a function of 
the distance $r$ to the crack tip. In a large system, an arbitrary
large crack can thus in principle produce an arbitrary large stress increment, which 
can thus overcome any rupture barrier created by the heterogeneities.
We thus expect the rupture to belong to a different class than for bounded
stress load transfer.

A rupture of size $a$ generates a stress intensity factor 
equal to $\sigma_{drop} \sqrt{a}$, where $\sigma_{drop}$ is
the stress drop. Let us call $\Delta \sigma$ the typical amplitude
of the stress barriers preventing the progression of cracks.
Two cases appear.
\begin{itemize}
\item $\sigma_{drop} \sqrt{a} < \Delta \sigma$\,: the amplitude of stress enhancement
 is smaller than the quenched
heterogeneity. The latter thus dominates and we expect an organization similar
to that observed in the previous case of limited load transfer.
\item $\sigma_{drop} \sqrt{a} > \Delta \sigma$\,: sufficiently large
cracks will always create stress transfer larger than the pre-existing
barriers. Beyond a characteristic nucleation size $a^*$ given by
$\sigma_{drop} \sqrt{a^*} \simeq \Delta \sigma$, cracks will not be stopped
and will always break through the system. 
\end{itemize}
This one-body argument seems to favor the idea that rupture in systems with
long-range elasticity should be ``first-order'', i.e. of the Griffith type.
As the deformation is building up, damage first grows progressively and one should witness 
an increasing cooperativity of crack growth and coalescence up to the critical 
size $a^*$, at which stage a different regime is switch on and
the macroscopic rupture is triggered abruptly. This is similar to the
situation occurring during the nucleation of a new phase in first-order phase
transitions in which droplet 
fluctuations below the critical radius suddenly leave place
to the unstable phase growth when a droplet reaches the critical size.

This scenario turns out to be wrong and the reason for this is that the ``mean-field'' argument
leading to the characteristic nucleation size $a^*$ neglects long-range 
interactions between many such characteristic cracks in a large system. Solvable models and
numerical 
evidence indicates that the rupture is in fact critical, i.e. exhibits an
increasing cooperativity of crack growth and coalescence up to the 
macroscopic rupture itself \cite{Herrmann,Roux,tfrefs,Zapperi}.
Ref.\cite{Zapperi} sees  breakdown in
disordered media as a first-order transition, while they present ample
evidence of scaling of the fraction of broken bonds and of the divergence of the
characteristic size up to the macroscopic 
rupture. Their claim is based on the fact that the fraction of broken bonds
has a discontinuity with a jump just at rupture. We think that this 
is confusing as it does not distinguish these results from the genuine abrupt behavior
of Griffith rupture, with no significant precursors and not divergence of
a characteristic length. A way to reconcile these points of view is to notice that 
critical phenomena in random systems have in general more than a single correlation
length. In rupture, one correlation length is the characteristic size of the damage, which
diverges upon appraoching the global rupture. A second charateristic length is the
size of the system that dictates the size of the macroscopic fracture.
We also stress that the power law scaling found in rupture in random media
does not correspond to some sort of ``spinodal'' analog of a first-order transition. 
Indeed, ``spinodal'' decomposition corresponds to the long-time coarsening of the phase
while rupture deals with the time behavior close to a critical control parameter value.
We also note that, in elastic-plastic transitions in heterogeneous systems, there
is not jump and the transition qualifies as critical \cite{plasticroux}. Since a model of
rupture in random media can be mapped onto this
elastic-plastic transition  \cite{Maloy}, this further substantiates the idea that
rupture in sufficiently heterogeneous media has the properties of a critical 
phenomenon.

These results have been obtained for sufficiently large disorder. The question arises
whether the critical nature of rupture in the presence of long-range elasticity
survives for all non-vanishing disorder or if rupture becomes truly abrupt for
sufficiently weak disorder. Indeed, we know that, for exactly zero disorder,
rupture is abrupt with no precursors or any diverging characteristic length\,: 
in fiber models of instance, identical fibers all break
simultaneously when the stress reaches their common threshold. 
Our results below show that the subtlety, richness and complexity of rupture processes in
heterogeneous media may be seen \cite{First} to arise from 
the non-commutation of the two limits $(q \to \infty, \Delta
\to 0)$, where $q$ is the order of the moment $\langle \sigma^q \rangle$
of the stress distribution and $\Delta$ is the amount of disorder,
in the same way that the non-commutativity of limits
is at the crux of some of the major outstanding problems in physics \cite{Berry} such
as turbulence (viscosity $\to 0$ ; time $\to \infty$) and quantum chaos ($h \to
0$ ; time $\to \infty$).
More practically, this corresponds to the non-commutation of the limits 
$(p \to p_r, \Delta \to 0)$, where $p_r -p$ is the parameter measuring 
the distance from rupture. $p$ can be the time, stress, or strain depending on 
the system and boundary conditions.
We now analyze carefully how the response of a system going to rupture depends 
on the amplitude of disorder.

\section{Model and Results}

We study the thermal fuse model \cite{tfrefs} which has recently received 
experimental scrutiny \cite{Lamai}. Let us recall briefly its definition that 
will be useful for the discussion below.
Fuses are put on the bonds of a square lattice of size $L \times L$. A fixed current $I$ is
imposed at time $t=0$ accross the two-dimensional lattice.
The fuses are heated by a generalized Joule effect
(electric power $\sim$ (current)$^b$) 
\begin{equation}
C{dT_n \over dt} = g_n^{-1} I_n^b ~, 
\label{ert}
\end{equation}
where $T_n$ is the temperature of the $n$th fuse, $C$ its specific heat, $g_n$ its
conductance, $I_n$ the current flowing in that fuse.
A fuse breaks down, becoming an insulator, when
its temperature $T_n$ reaches a given threshold, the same for all. The heterogeneity 
is on the conductances $g_n$ of the fuses, distributed according to a uniform
distribution in the interval $[1-\Delta/2 ;1+\Delta /2 ]$.
$\Delta$ is the measure of the amplitude of disorder and varies from $0$ (no
disorder) to $2$ (maximum disorder). As a result
of the delay and relaxation effects embodied in the dynamics of the heating
of each fuse, a wealth of novel
behaviors emerge, with fractal cracks and critical
behavior in the time domain with exponents being continuous functions of $b$
\cite{tfrefs}. The motivation behind this model was to introduce the
simplest genuine dynamical rupture model which is still amenable to 
quasi-static elastic calculations. This is done by coupling the dynamical
variable to the elastic fields only at the time of each rupture event.
This model has the following equivalent mechanical formulation in mode III
(antiplane) elasticity. Currents
become forces, potentials become displacements, the temperature of a fuse
becomes the damage variable of this element, which when reaching one triggers
its breakdown.

In previous investigations \cite{tfrefs,Lamai}, it was shown that the electric resistivity of
the system diverges upon approaching the global rupture occurring at time $t_r$ as
\begin{equation}
R \sim (t_r -t)^{-\alpha} ~,
\label{eq1}
\end{equation}
with an exponent $1 \leq \alpha(b) \leq 2.3$ in 2D as $b$ decreases from $+\infty$ to $0$.
It was also observed qualitatively that this power law is observed in a 
``critical region'' that shrinked as the disorder amplitude $\Delta$ decreases.
The question is whether the critical region vanishes at a finite non-zero value
of $\Delta$ or just shrinks continuously with $\Delta$, and if so, how?

We study the resistivity $R_{L,\Delta}(t)$ as the observable
defined for a system of size $L$ with disorder amplitude $\Delta$ at time $t$. 
Building on the critical rupture hypothesis, we test whether $R_{L,\Delta}(t)$ can
be represented as
homogeneous function. From the $\Pi$ theorem for homogeneous functions
\cite{barenblatt}, we write 
\begin{equation}
R_{L,\Delta}(t) =
[{\tau \over \Delta ^\delta} ]^{-\alpha} 
G \bigg( {\tau \over \Delta ^\delta}, \tau L^{1 \over \nu}  \biggl)
\label{eq2}
\end{equation}
where $\tau \equiv {t_r -t \over t_r}$. The scaling function $G(x,y)$ should 
scale as 
\begin{itemize}
\item 
$L^{-1/\nu} << \tau << \Delta^{\delta}$ (corresponding to $t \to t_r$ with
$\Delta$ fixed and finite size effects not important). This leads to
$G(x \to 0, y \to \infty) \to constant$, i.e. $R 
\sim [{\tau \over \Delta ^\delta} ]^{-\alpha}$;
\item  
$L^{-1/\nu} << \tau$ and $\Delta^{\delta} \leq \tau$
(corresponding to $\Delta \to 0$ with
$t$ fixed and finite size effects not important). This leads to
$G(x \to \infty, y \to \infty) \to x^{\alpha}$ , i.e.
$R \sim constant$;
\item 
$\Delta^{\delta} \leq \tau \leq L^{-1/\nu}$ (corresponding to $\Delta \to 0$ with
$t$ fixed and finite size effects important). This leads to 
$G(x \to \infty, y \to 0) \to y^{\alpha}$ , i.e.
$R \sim L^{\alpha \over \nu}$.
\end{itemize}
We have introduced the usual finite size scaling ansatz with a characteristic width of the
transition scaling as $L^{-1/\nu}$, where $\nu$ is the correlation length
exponent. We have also assumed that 
the disorder amplitude $\Delta$ determines the size of the region over which 
the fracturing stays critical and have introduced the disorder exponent $\delta$ which may
a priori depend on $b$.

The finite size scaling with the system size $L$ has been previously tested successfully
\cite{Herrmann,Roux,tfrefs,Zapperi}. We turn to a test of Eq~(\ref{eq2}) with
respect to the disorder dependence. Figures 1 and 2 show
 the scaling function $G$ defined from Eq~(\ref{eq2})  as a function of $x$
for two different values of the exponent 
$b$ with various values of the disorder field $\Delta$. The  
data points are obtained by sampling over 25 independent simulations on 
a square lattice of size $180 \times 180$ tilted at $45^\circ$, using the
method of \cite{tfrefs}. These results thus
correspond to a significant computational effort.
The exponent $\alpha$ was chosen $\alpha (0.5) =0.9$ and $\alpha (2) =0.3$ 
respectively, in accordance to the power law fits done previously \cite{tfrefs}. 
We found that data collapse was optimal using $\delta =1$ for both 
values of $\alpha$.  

Figures 2, for $\alpha (b=2) =0.3$, show an excellent verification of the scaling relation 
Eq~(\ref{eq2}) with an almost 
complete overlap for the various values of disorder $\Delta$ from $0.1$ to $0.8$. 
As expected  
$G(x) \propto x^{\alpha}$ for large $x$ (large $t_r-t$), and crosses over 
to a constant for small $x$ ($t$ close to $t_r$). The data for the largest
disorder $\Delta = 1.6$ departs from the master curve\,: this can be attributed to the fact that 
the resistivity is modified by the pre-existing heterogeneity, even in absence of damage. 

For $\alpha (b=0.5) =0.9$, simulations with different disorder 
collapse neatly onto one scaling curve for $x > 5~10^{-3}$, going to the expected
powerlaw $G(x) \propto x^{\alpha}$ for the largest $x$ and 
a constant in the region $5~10^{-3} \leq x \leq 2~10^{-2}$. For smaller values of $x$,
the condition $L^{-1/\nu} << \tau << \Delta^{\delta}$, for which $G$ is a constant
and finite size effects are not felt, is not obeyed anymore and 
is replaced by $L^{-1/\nu} << \tau << \Delta^{\delta}$, for which 
$g(x \to 0, y \to 0) \to y^{\alpha} \propto x^{\alpha}$. We thus expect and observe
indeed that the scaling function $G$ should return to a power law for the smallest values
of $x$ after the intermediate plateau. This finite size effect is all the strongest
for the smallest $b$, since smaller  
values of $b$ give a more diffuse network of cracks \cite{tfrefs}, whose fractal dimension
increases with $b$, from  the value $1$
up to the percolation cluster dimension $\approx 1.9$ in 2D. As for figure 2, the data for
the largest disorder $\Delta = 1.6$ departs from the master curve, for the same reason.

There is a simple explanation for the universal value of the disorder exponent $\delta = 1$, which
does not seem to depend on $b$ while the other exponents (the resistivity exponent $\alpha$ and the 
crack fractal dimensions) do \cite{First}. This has to do with dimensional analysis of (\ref{ert}).
Indeed, (\ref{ert}) is a scale invariant equation and since the electric-elastic equations
are also scale invariant, 
dimensional analysis should give the correct scaling law in the time domain. Eq. ~(\ref{ert})
indicates that the time scales linearly with $g$. Therefore, the spread in time scales scale with the
width of the distribution of $g$. This leads to the prediction that the width of the critical 
domain for rupture must scale linearly with the disorder amplitude $\Delta$, hence $\delta = 1$, 
irrespective of the value of the exponent $b$. A similar argument gives that
 $t_r \sim I_{tot}^{-b}$ \cite{tfrefs}, where $I_{tot}$ is the total
current applied to the system. 
As already pointed out, the thermal fuse model exhibits a simplification
 in the coupling between the damage field (temperature)
 and the elastic field (electric current), which occurs only at the time of fuse breaking. This is the 
 origin of this remarkable universal scaling law as a function of the disorder.

\section{Conclusion}

We have shown that, in a simple dynamical model of 
rupture in heterogeneous media incorporating long-range elastic forces,
scaling holds as a function of the amplitude of the disorder\,: the width of the 
critical region is linearly proportional to the amplitude of the disorder. As a consequence,
we conclude that, at least in this prototype model, dynamical rupture in random media is a genuine
critical phenomenon that occurs for any non-zero disorder however small. In pratice however, the
region of control parameter over which the critical regime can be observed becomes
exceedingly small for small disorders and may thus mislead to the conclusion of an abrupt 
behavior more similar to a first-order transition. This situation is different
from the one of bounded load transfer previously investigated \cite{First} in which there
is a finite disorder amplitude below which the critical regime disappears and is replaced
by an abrupt first-order regime. It would be worthwhile to test these behaviors on other
models and in experiments to confirm and extent the present classification.

\vspace{0.5in}
We are grateful to C. Vanneste for his help in the numerical simulations.
D.S is supported by NSF under grant EAR9615357 and EAR9706488.
J.V.A wishes to acknowledge support from
the European Union Human Capital and Mobility Program
contract number ERBCHBGCT920041 under the direction of Prof. E. Aifantis.


\pagebreak

\begin{figure}
\begin{center}
\input{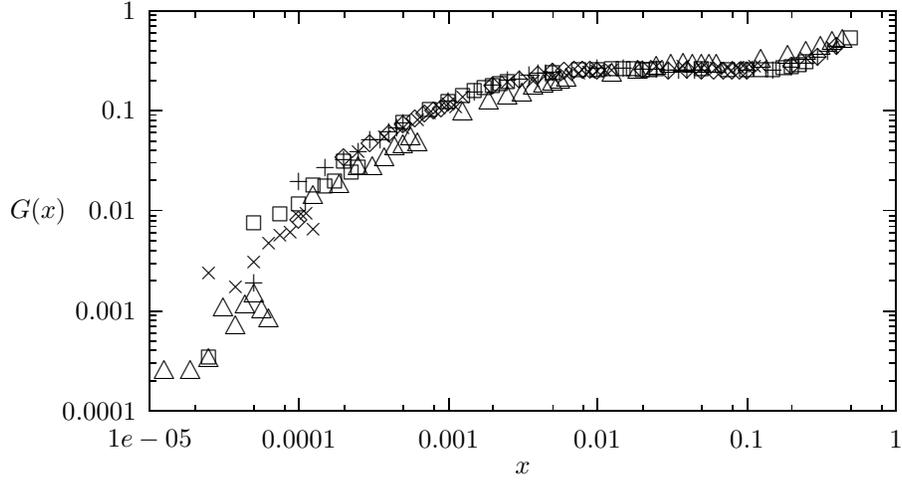}
\caption{Scaling function $G(x)=R(x,y) ~x^{\alpha (b)}$ versus $x$ for $b=0.5$, using 
$\alpha (0.5) =0.9$ as obtained independently from \protect\cite{tfrefs}. 
Disorder amplitude $\Delta  =$ $0.1 (\diamond )$, 
$0.2 (+)$, $0.4 \Box$, $0.8 (\times )$, and $1.6 (\triangle )$. 
The best overlap was obtained using $\delta =1$.}
\end{center}
\label{scalingplot1}
\end{figure}

\begin{figure}
\begin{center}
\input{tfscalingfig2.tex}
\caption{Scaling function $G(x)=R(x,y) ~x^{\alpha (b)}$ versus $x$ for $b=2$, using 
$\alpha (2) =0.3$ as obtained independently from \protect\cite{tfrefs}.
Disorder amplitude $\Delta  =$ $0.1 (\diamond )$, 
$0.2 (+)$, $0.4 \Box$, $0.8 (\times )$, and $1.6 (\triangle )$.  
The best overlap was obtained using $\delta =1$.}
\end{center}
\label{scalingplot2}
\end{figure}

\end{document}